\documentclass[conference]{IEEEtran}
\usepackage{balance}

\usepackage{shortcuts}

\begin{document}
%
\title{Localizing Patch Points From One Exploit}

\author{
\IEEEauthorblockN{Shiqi Shen \quad Aashish Kolluri \quad Zhen Dong \quad Prateek Saxena \quad Abhik Roychoudhury}
\IEEEauthorblockA{{\em Computer Science Department}, {\em School of Computing}\\ 
  {\em National University of Singapore}\\
  \{shiqi04, aashish7, zhen.dong, prateeks, abhik\}@comp.nus.edu.sg}}



\maketitle



\begin{abstract}

Automatic patch generation can significantly reduce the window of
exposure after a vulnerability is disclosed. Towards this goal,
a long-standing problem has been that of patch {\em localization}: to
find a program point at which a patch can be synthesized. We present
\tool, one of the first systems which automatically identifies such a
location in a vulnerable binary, given just one exploit, with high
accuracy. \tool does not make any assumptions about the availability
of source code, test suites, or specialized knowledge of the
vulnerability. \tool pinpoints valid patch locations in large
real-world applications with high accuracy for about \successratio of
\fuzznum CVEs we study. These results stem from a novel approach to
automatically synthesizing a test-suite which enables
probabilistically ranking and effectively differentiating between
candidate program patch locations.
 
\end{abstract}

%

\section{Introduction}

Security vulnerabilities can remain unpatched for months after their initial
disclosure~\cite{38dayspatch, patchisharder}.  This has motivated recent
interest in automatic patch generation techniques for security
vulnerabilities~\cite{senx}, and more broadly, general program repair
techniques have independently advanced as well~\cite{CACM-roychoudhury}. These
techniques recommend deployable patches, which can be inspected and finalized
by the developer, reducing the time window of exposure to exploits.

The problem of automatic patching can be decomposed into two: patch
localization and patch synthesis. {\em Patch localization} aims to identify
particular locations in the program at which one can patch the program. Patch
synthesis aims to generate a fix (or program statements) modifying the program
variables at the identified locations to eliminate the vulnerability. It has
been shown that once a suitable location is identified, synthesis is feasible
through a number of prior techniques~\cite{CACM-roychoudhury,assiri2016flForAPR}. In this
paper, we focus on the patch localization problem, which has remained an
elusive practical goal.

User experience studies highlight that developers expect an automated
patch localization tool to be fairly accurate.  Developers are willing
to inspect a handful (typically 5) patch locations recommended by
localization tools~\cite{kochhar2016practitioners}.  However, prior
techniques yield recommendations with such accuracy in less than $50\%$ of the
programs studied and most techniques do not scale to large
applications~\cite{kochhar2016practitioners}. Furthermore, these
techniques make many assumptions, such as the availability of program
source code, a plethora of auxiliary code metrics beyond just source
code~\cite{li2019deepfl}, or extensive test-suites~\cite{TarantulaICSE02}.

In this work, we tackle patch localization for vulnerabilities in
real-world programs. We make localizing fixes possible with minimal
assumptions. We assume that the only available information is a single
executable exploit, such as those typically available in CVE reports,
and the program itself. Our technique works with program binaries and
does not assume access to source code. Our techniques are agnostic to
the type of vulnerability being patched---we only assume that an
observable outcome (e.g. a crash) can characterize exploiting runs.
Our techniques do not assume access to an external test-suite.  Lastly,
our technique is based on runtime instrumentation and observing
recorded program states; we carefully avoid using any sophisticated
program analyses which may not scale on binaries.
\update{While the suggested locations are eventually given to a human
developer, we believe that working with minimal assumptions makes our
technique more deployable in any stage of the development pipeline.}
One can also view our proposed technique as the first part of a larger
system that automatically synthesizes full patches at the binary
level, though we only focus on the localization step in this work.

Our general approach is based on statistical fault localization~\cite{Survey}.
 Program locations can be assigned a probability estimate or score, which
measures how likely is the program going to be exploited if  the instructions
corresponding to those locations are executed. High scores suggest that it is
both sufficient and necessary for these instructions to be executed in order
to reach the vulnerability and exploit it. \update{Patching at these locations is
likely to eliminate the vulnerability while minimizing the impact on benign
program behaviors---we explain why this is so in
Section~\ref{sec:statframework}.}

Though statistical fault localization has been proposed about two
decades ago, it has not produced high-fidelity results so far.  One
question has remained unresolved in prior works: {\em under which
  input distributions should these probabilistic scores be estimated?}
This issue is of fundamental importance to statistical localization
because the probabilistic quantities of interest can not be robustly
estimated under arbitrary input distributions.  Test inputs not
designed with the objective of patch localization will often lead to
{\em over-fitting}. The estimated scores will either be biased by
failing (exploit) tests or benign ones, depending on which of these
dominate the provided test-suite. \update{Ad-hoc or poorly chosen test-suites
fail to distinguish between candidate program locations, essentially
assigning them the same score. The problem is acute in large
applications in which more than a thousand locations may be
observed in single program execution.}
To avoid over-fitting, the key problem reduces to synthesizing a
specific kind of test-suite which we refer to as {\em concentrated}. A
concentrated test-suite exercises a sufficient diversity and quantity
of program paths in the neighborhood the exploit path.  The
probabilistic quantities of interest can be estimated robustly using
such a test-suite.

We propose a simple procedure to construct a concentrated test-suite
from a single exploit. The procedure is a new form of directed
fuzzing, which we call as {\em concentrated fuzzing} or
\fuzzer. \update{Directed fuzzing techniques have witnessed rapid
  advances recently, however, their prime application has been for
  crash reproduction and patch testing~\cite{AFLGo}.} Our work is the
first, to the best of our knowledge, to propose its application for
patch localization. \update{\fuzzer is unlike other forms of directed
  fuzzing, which primarily aim to reach a particular location.  The
  goal of \fuzzer is to estimate the probability of each branch being
  executed in exploiting and benign runs. It generates inputs that
  exercise paths in the neighborhood of the path taken under the given
  exploit to estimate these probabilities. \fuzzer is simple to
  implement. It requires instrumentation of a small number of program
  points.} We implement our proposed techniques in a tool called
\tool.

Our main empirical finding is that our new approach scales to large
applications and has high accuracy in finding patch
locations. \shiqi{We evaluate \tool on \fuzznum CVEs on programs
  ranging from $10$K - $2$M LOC. We use a single exploit available
  from the CVE report and the program binary for each benchmark.} In
about \successratio of these benchmarks, \tool pinpoints at least one
location in its Top-5 ranked outputs, where a patch equivalent to the
{\em developer-generated} patch exists. Our techniques can be used to
localize from crashes, exploits, or any other oracle of failure---our
tested CVEs cover many common types of security vulnerabilities. The
locations highlighted by \tool can be passed to a patch synthesis
tool; in our evaluation, we manually synthesize the patch to confirm
that the vulnerability can be patched at the highlighted locations.

\paragraph{Contributions.}
We make the following contributions:

\begin{enumerate}
	\item We present \tool, a patch localization tool with very
          few assumptions. It takes a vulnerable application binary and an exploiting test case
          as input and outputs Top-5 candidate locations.

	\item \tool is the first work to propose directed fuzzing for
          localization. We propose a novel directed fuzzing technique
          called concentrated fuzzing (\fuzzer), which explores benign
          and exploiting paths in sufficient diversity around the
          given exploit path. \update{We demonstrate that the
            test-suites generated using \fuzzer are able to avoid
            over-fitting to specific test types and allow
            distinguishing effectively between different candidate
            locations.}

	\item We evaluate the efficacy of \tool on \fuzznum CVEs in
          large real-world applications.  \update{\tool localizes
                    patches for about \successratio of the CVEs within Top-5
                    locations in around $4$ hours per CVE.}
\end{enumerate}

\section{Motivation \& Problem}~\label{sec:prob}

Automated patching consists of two steps: patch localization and patch
synthesis. If accurate patch locations can be recovered, existing
program synthesis tools can be used to generate vulnerability
patches~\cite{CACM-roychoudhury}.
We focus on the patch localization problem in this work.

\subsection{Problem}

To keep the assumptions minimal, we only assume access to a vulnerable
application binary $Prog$, an exploit input $i_e$ and the corresponding
security specification. The specification is violated while executing $Prog$
with $i_e$.  Given the specification, we implement a {\em vulnerability
oracle} to detect whether $Prog$ gets exploited under a specific test-case. A
program crash can be an oracle for memory safety. For numerical errors (e.g.,
divide-by-zero), the processor provides hardware bits that can be checked. No
source-level information is necessary to implement the oracle.

\begin{figure}[t]
	\begin{lstlisting}[style=JavaScript, language=C, xleftmargin=0.3cm, captionpos=b]
	   static int PixarLogDecode(TIFF* tif){
	   	...
		+  /* Check that we will not fill more than what was allocated */
		+	if (sp->stream.avail_out > sp->tbuf_size){
		+		TIFFErrorExt(tif->tif_clientdata, module, "sp->stream.avail_out > sp->tbuf_size");
		+		return (0);}
			// write into sp->stream
		    int state = inflate(&sp->stream, Z_PARTIAL_FLUSH);
		}
		// Function that cleans up pixarlog state
		static int PixarLogCleanup(TIFF* tif){
			...
			_TIFFfree(ptr);
		}
	\end{lstlisting}
	\caption{Developer provided patch for CVE-2016-5314, Patch is not at the crash location.}
	\label{fig:20165314}
\end{figure}

For the rest of the paper, we use the notion of {\em observed branch
locations}. Concretely, when we execute the program binary with an input, we
observe only the branch locations executed. Based on which branches are
observed in the execution of specific inputs (generated via fuzzing), certain
branches are predicted as patch locations by our technique. \update{The
localization is performed at the basic block level, since we want \tool to
work even when the source code or debug symbols are unavailable. We say the
prediction is correct, if the basic blocks immediately preceding/succeeding
the predicted branch can be modified (or are modified in the developer-
generated patch) to fix the vulnerability}. \ash{Note that the final patch, whether
manually written or automatically generated, can be applied anywhere inside
the predicted basic blocks.}

For security vulnerabilities, one could assume that applying a patch
right before the crash location is sufficient~\cite{senx}. This is not
a valid assumption, and in fact, developer-provided patches often do
not follow such a pattern (see our Section~\ref{sec:eval}).
The task of identifying a correct patch location is more subtle. There
are two objectives for a correct patch: 1) stopping the failing (or
exploiting) program runs, and 2) preserving compatibility with benign
runs.  \ash{Consider a bug in the open-source library,
  LibTIFF~\cite{libtiff}, shown in Figure~\ref{fig:20165314}. It is a
  heap overflow vulnerability involving the buffer
  \texttt{sp->stream}.  It arises from out-of-bound writes in
  \texttt{PixarLogDecode} function at Line $8$ without checking the
  buffer length, which causes the head of the next heap to be filled
  with arbitrary data.  The crash occurs when the invalid pointer is
  freed in another utility function called \texttt{\_TIFFfree}. Since
  it is a utility function which is used by other functions (such as
  \texttt{PixarLogCleanup}), it is {\em not} known whether the pointer
  to be freed is invalid or not at the crash location. Moreover, any
  changes to the utility function \texttt{\_TIFFfree} would change
  other benign behaviours of the program.} Instead, a better way to
patch this program is to prevent the out-of-bounds write since the
buffer size is known during the write---as done in the developer
provided patch. Notice that it requires domain-specific knowledge to
infer the correct patch location and the complexity of this inference
will vary with each bug.

\subsection{A Statistical Framework}\label{sec:statframework}

Consider the execution of $Prog$ under $i_e$. Let the set of program branches
encountered in the execution before violating the security specification be
$V= \{v_1, v_2, \cdots, v_n\}$ and the execution trace be $U=\langle u_1, u_2,
\cdots, u_m\rangle$. Each $u_i \in U$ is an instance of a branch $L(u_i)$
where $L(u_i) \in V$. Now, consider the execution of $Prog$ under a wide
variety of inputs $I$, possibly generated by fuzzing with the exploiting input
as the seed. Under each input, we see a subsequence of $U$ in the execution
trace with a subset of $V$ observed.  Each $v_i$ and $u_j$ are associated with
a Bernoulli random variable $X_i$ and $Y_j$ respectively. Each random variable
takes a value $1$ if it is observed in that execution trace, else $0$. 
Similarly, the violation of the security specification can be captured by a
Bernoulli random variable $C$, which is $1$ if the program violates the
security specification, and $0$ otherwise.

This abstraction allows us to reason about the statistical correlation between
the events where $C$ and $X$ take on certain values. If an event $X_i$=$1$
happens in {\em all} of the exploit traces in the input set $I$, one could
induce that $(C$=$1) \Rightarrow (X_i$=$1)$, which indicates that patching at
$v_i$ might avoid all exploits seen. However, it is possible that the patch
might significantly change the benign behavior of the program. Conversely,
consider an event such that whenever it occurs the program gets exploited. One
can then induce that $(X_i$=$1) \Rightarrow (C$=$1)$. Since this event is not
observed on any benign test, patching at $v_i$ is likely to have the least
impact on benign behaviors, with the caveat though that such a patch may not
cover {\em all} exploits seen. The best patch should prevent all exploits
while having the least impact on benign runs. Overall, an event which is both
necessary and sufficient carries a strong signal of the root cause underlying
the exploit and is an ideal patch location candidate.

Consider the branch $v_i$ such that the instances of $v_i$ appear as $U_i =
\{u_j| L(u_j) = v_i\}$.  \update{As a result, we can compute the probability of each
branch location being witnessed as the probability that at least one of its
instances is witnessed:}
\[
P(X_i=1) =   P(  \bigcup_{u_j \in U_i}  Y_j=1)
\]
\update{We now compute two scores for each branch location $v_i$:}
\begin{itemize}
\item Necessity score is $P(X_i=1|C=1)$, the likelihood of observing at least one instance of the branch on an exploiting input;
\item Sufficiency score is $P(C=1|X_i=1)$, the likelihood of getting exploited on an input where at least one instance of the branch is observed.
\end{itemize}

Similar to the scores for each branch location, we can also define necessity
and sufficiency scores of a single branch instance $u_j$ which are $P(Y_j=1
|C=1)$ and $P(C=1|Y_j=1)$ respectively. Branches with the highest $K$
necessity and sufficiency scores are highlighted to the developer. \update{The
developer can then synthesize a fix in or around these locations.}

\subsection{Our Approach}\label{sec:approach}

The framework presented thus far is similar to the underpinning of a long line
of works on statistical fault isolation~\cite{Survey,TarantulaICSE02}.
However, there is a central issue left unaddressed, which we study here: {\em
under which input distribution should the probabilities be estimated?}

Consider computing the necessity score for each instance $P(Y_j=1|C=1)$. As
the probability of observing $u_j$ deep down in the execution of an exploit
may be very small, \update{the probability estimates will {\em over-fit} the given
test-suite if the test-suite only contains a few observations over $u_j$.} The
same phenomenon arises when computing sufficiency score, as most instances
with very few (or no) observations in the benign runs, leading to an
artificially high score. In other words, an arbitrary test-suite is unlikely
to {\em distinguish} between $u_j$ appearing in the exploit trace for patch
localization.

\paragraph{Need for Sufficient Observations of $Y_j=1/0$.}
The crux of our problem is to construct a test-suite which distinguish $u_j$
with only an exploit input. To explain it, we factorize the necessity score
into:
\begin{equation} \label{eq:eq1}
\small
\begin{split}
P(Y_j=1|C=1) &= P_1 \times P_2 + P_3 \times P_4 \\
P_1 &= P(Y_j = 1| C=1, Y_{j-1}=1) \\
P_2 &= P(Y_{j-1}=1|C=1) \\
P_3 &= P(Y_j = 1| C=1, Y_{j-1}=0) \\
P_4 &= P(Y_{j-1}=0|C=1)
\end{split}
\end{equation}
The terms $P_1$ and $P_3$ differentiate the correlation of $u_j$ to $C$ and of
$u_{j-1}$ to $C$. A high ratio of $P_1$ to $P_3$ means that $u_j$ and
$u_{j-1}$ are equally correlated to $C=1$, as the exploit inputs always see
the co-occurrence of $u_j$ and $u_{j-1}$. However, a low ratio means that
$u_j$ is more likely to be observed when $C=1$ than $u_{j-1}$, hence
distinguishing them. It is thus important to construct a test-suite from which
terms $P_1$ and $P_3$ can be estimated robustly. In particular, we need a
test-suite with a sufficiently large number of exploit traces observing
$u_{j-1}$ and not observing $u_{j-1}$ respectively.

This motivates the need for what we call a {\em concentrated} test-suite: a
test-suite that has sufficiently many tests, both observing and not observing
each $u_j$. The concentrated test-suite would explore paths in the
neighborhood of the exploit trace. This is apparent in $P_1$ in
Equation~\ref{eq:eq1}: it is the probability of observing $u_j$, conditioned
on the fact that we have observed $u_{j-1}$. This can be seen as following the
trace of the given exploiting input $i_e$ up to $u_{j-1}$ but not necessarily
following the exploit trace after $u_{j-1}$.

To generate a concentrated test-suite, we propose a new form of directed
fuzzing technique called {\em concentrated fuzzing}. Concentrated fuzzing is
constructed in a principled way, and it carefully tries to avoid artificially
biasing test cases towards observing the events (i.e., $Y_j=1$ and $C=1$)
which we will estimate from. This highlights the importance of having various
kinds of test cases including the exploits, the benign cases, the cases
reaching the crash location and the cases deviating from the crash location.
From a test-suite created by concentrated fuzzing, we localize the patches and
show the corresponding empirical results (see Section~\ref{sec:eval}).

\paragraph{Remark}.
Our work is the first to utilize directed fuzzing as a solution for patch
localization. While heavy-weight alternatives such as those based on symbolic
execution are possible solutions, these face challenges in scaling on
binaries~\cite{jin2013f3}. However, fuzzing is simpler to implement and
scales. Other forms of fuzzing (e.g., AFL~\cite{afl} and Honggfuzz~\cite{honggfuzz}) optimize for different objectives
from concentrated fuzzing.  Their goal is to cover more program paths, whereas
our goal is to explore only the neighborhood of a given exploit trace to find
likely patch location of the exploitable vulnerability. \update{We also compare our
test-suite from concentrated fuzzing with directed fuzzing tools like AFLGo
\cite{AFLGo} in our empirical evaluation.}

\section{Concentrated Fuzzing}
\label{sec:design}

\begin{algorithm}[t]
\caption{\update{Meta algorithm for concentrated fuzzing.}}\label{alg:ideal}
\begin{algorithmic}[1]
\State \textbf{Input: } Exploit input $i_e$, Execution trace $U$, Instrumented binary $Prog$
\State \textbf{Result: } Test-suite $T$
\State $T \leftarrow \emptyset$
\For{\textbf{each} $u_j \in U$}
	\State executeTillPrefix($i_e$, $u_j$, $Prog$)
	\For{$k$ from $1$ to $\alpha$} 
		\State $i_m$ = mutate($i_e$, $u_j$);
		\State $t_m$ = execute($i_m$, $Prog$); 
		\State $T \leftarrow T \cup \{t_m\}$
	\EndFor
\EndFor
\end{algorithmic}
\end{algorithm}

Constructing a concentrated test-suite is not straight-forward since the
probability of reaching $u_j$ which is deep inside the program is very low
under most input distributions. To achieve this goal, we propose a new
solution called  {\em concentrated fuzzing} (\fuzzer).

Algorithm~\ref{alg:ideal} shows the high-level algorithmic sketch of \fuzzer.
It creates a set of inputs fully exploring each $u_j$ (loop at Line $4$)
starting from $u_1$. In each iteration, the basic idea is to force the program
execution to reach $u_{j-1}$ (at Line $5$), and then generate sufficiently
many test cases that either reach $u_j$ (stay on exploit path) or not (loop at
Line $6$). \update{To ensure the reachability of $u_{j-1}$, many different strategies
can be used---for instance, we could pick a sample from the set of test cases
generated so far on which $u_{j-1}$ was observed, and replay the execution}.
For simplicity, we run the program with the given exploit $i_e$ to execute a
{\em prefix} of the exploit trace upto $u_{j-1}$. Then, we mutate $i_e$ to
generate test cases, some of which will observe $u_j$. Notice that if these
mutations are made arbitrarily, the execution may diverge off the prefix
early, failing to observe $u_{j-1}$. Section~\ref{sec:arch} describes the
details of a directed fuzzing approach where certain input bytes in $i_e$
remain unchanged, such that the prefix up to $u_{j-1}$ will be executed with
high probability. Random mutations to other bytes are created to
create sufficient many samples over the events $Y_j =1$ and $Y_j=0$.

\begin{figure}[t]
\centering
\begin{lstlisting}[style=JavaScript, language=C, xleftmargin=0.3cm, captionpos=b]
void write(int size, char *writeArr){
	for(int i=0; i<size; i++){
		writeArr[i] = "A"; // <---- buffer overflow
	}}

int write_array(int wsize, int msize){
	char *writeArray;
	if (wsize > 20)
		return -1;
	else {...}
	if (msize <= 10)
		writeArray = (char *)malloc(msize);
	else
		writeArray = (char *)malloc(2*msize);
	write(wsize, writeArray);// write wsize characters into writeArray
	...}

int main(int argc, char **argv){
	int a, c, tag;	
	...
	FILE *fp = fopen("input.txt", "r+"); // Read inputs from a file
	fscanf(fp, "%d, %d, %d", &a, &c, &tag);
	...
	if(tag == 1){ // Read from input file
		read_array(c, fp);
		...
		write_array(c, a);
	} else if(tag == 2){ // Write array
		write_array(c, a);
	} else if(tag == 3){ // split the input file
		...
		char array[10];
		write(10, array);
	}
	...
	return 0;}
\end{lstlisting}
\caption{
Code snippet for illustrating our fuzzing and ranking. There is a buffer
overflow in function \texttt{write}.
}
\label{fig:design_example}
\end{figure} 
\begin{figure}[t]
\centering
\includegraphics[width=1.0\linewidth]{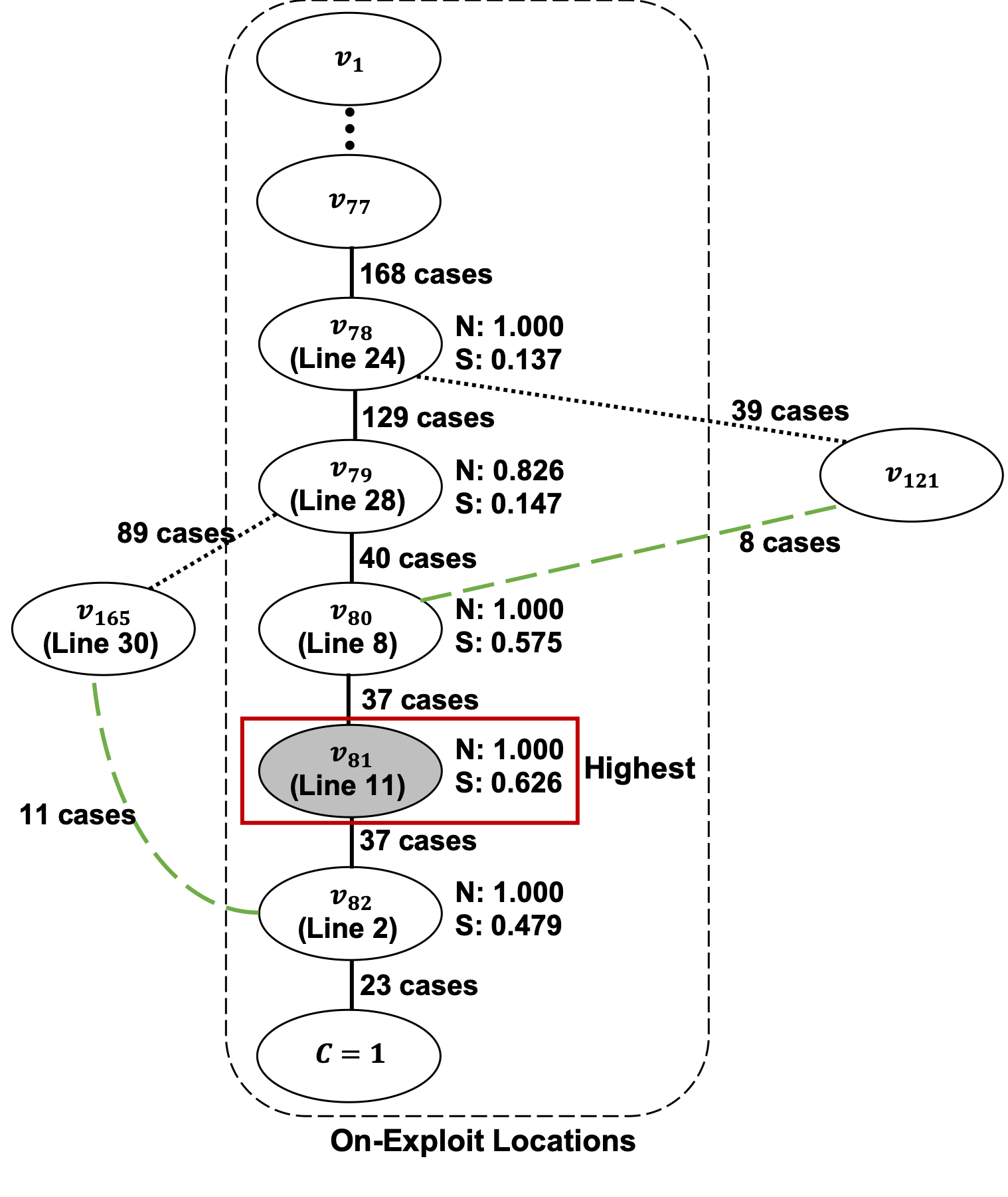}
\caption{
An example of concentrated test suite generated for the the code snippet in
Figure~\ref{fig:design_example} and its corresponding necessity (N) and
sufficiency (S) score. 
}
\label{fig:example_cl}
\end{figure}

\paragraph{An Illustrative Example.}
Figure~\ref{fig:design_example} shows a hypothetical vulnerable program which
has a buffer overflow in \texttt{write} function (at Line $3$). Let us assume
that the exploit input available is $i_e$=$(10,15,2)$ where the three bytes of
$i_e$ are inputs to the program and are read from an input file ``input.txt''
at Line $21$. On execution of $i_e$, the function \texttt{write\_array} is
invoked with the arguments \texttt{wsize}=$15$ and \texttt{msize}=$10$.  It is
easy to check that an additional check \texttt{msize $\geq$ wsize} on the size
of memory being allocated to the buffer \texttt{writeArray} at Line $11$ is an
ideal patch candidate. Notice that the patch is not close to the point of the
buffer overflow (Line $3$), where the variable \texttt{msize} is not even in
scope. Ideal location to patch can be far off from the exploit point, as in
this example.

To localize at the right patch location, \fuzzer generates a
concentrated test-suite to explore each on-exploit branch instance
$u_j$. Recall that the goal of patch localization is to identify the
right location $v_i$ for fixing the vulnerability rather than the
instance $u_j$. Thus, in Figure~\ref{fig:example_cl}, we summarize the
number of test cases generated for each $v_i$ in the concentrated
test-suite. Notice that there are many test cases for both observing
and not observing each on-exploit location. These tests are not
generated with the objective of causing an exploit. Biasing towards
following $v_i, v_{i+1}, \cdots, v_{n}$ would skew the samples.
Over-fitting to exploits and its effects are shown experimentally in
Section~\ref{sec:overfit}. This is an important difference to recent
works which explicitly aim to follow the path {\em suffix} of the
exploit after $v_i$~\cite{jin2013f3,robetaler2012isolating}. Our tests
seek to follow the given exploit path {\em prefix} up to $v_i$ and
then diverge.

Under the exploit $i_e$, the vulnerable program executes the branch at  line
$24$ ($v_{78}$), $28$ ($v_{79}$),  $8$ ($v_{80}$), $11$ ($v_{81}$) and $2$
($v_{82}$) which are on-exploit locations. Taking $v_{80}$ as an example,
\fuzzer forces the execution of the program to observe branches $\{v_1,
\cdots, v_{79}\}$ and gets $48$ cases observing $v_{80}$ and $89$ cases that
do not observe $V_{80}$ (but instead observes $v_{165}$). Among these $48$,
$23$ cases trigger the buffer overflow while the remaining $25$ cases do not.
\fuzzer generates sufficient test cases for each $v_i$ in this way and finally
computes the necessity and sufficiency score. As shown in
Figure~\ref{fig:example_cl}, $v_{81}$ has the highest scores compared with the
other on-exploit locations, which is the exact location of the ideal patch.

Careful readers may notice that our procedure executes $i_e$ to force the
prefix up to $u_{j-1}$, when generating the concentrated test cases for $u_j$.
This corresponds to the conditional probability $P(Y_j=1|C=1, Y_1=1, \cdots,
Y_{j-1}=1)$, as opposed to $P(Y_j=1|C=1, Y_{j-1}=1)$ desired in
Section~\ref{sec:prob}. Notice that the latter is a marginal probability which
is a summation over exponentially many conditional probabilities:
\begin{gather*}
\small
\sum\limits_{q_1 \in\{0,1\}}\cdots\sum\limits_{q_{j-2} \in \{0, 1\}} \\
P(Y_j=1|T_1=q_1,\cdots, T_{j-2}=q_{j-2},C=1, T_{j-1}=1) \\
\times P(T_1=q_1,\cdots, T_{j-2}=q_{j-2}|C=1, T_{j-1}=1)
\end{gather*}

Since measuring the marginal would require sampling across all paths leading
up to $u_j$, estimating it may require intractably many test cases. However,
as our experiments demonstrate, the $P(Y_j=1|C=1, Y_1=1, \cdots, Y_{j-1}=1)$
serves as a good proxy for the marginal we desire. To understand why such a
proxy works well in practice, let us consider the situation where there is
conditional independence: the probability of observing an event, conditioned
on having reached the point which observes $u_{j-1}$, is independent of the
probability of reaching $u_{j-1}$ in the first place. In such a case,
$P(Y_j=1|C=1, Y_1=1, \cdots, Y_{j-1}=1)$=$P(Y_j=1|C=1,Y_{j-1}=1)$. In our
example, the probability of reaching Line $11$ is indeed almost independent of
the probability of causing exploits---for any given value of \texttt{msize},
there are many values of \texttt{wsize} that would lead to exploits.
Intuitively, when the vulnerability is dependent on a small number of
branches, such ``locality'' creates conditional independence of the form that
our technique works well with.

\begin{figure}[t]
\centering
\resizebox{0.42\textwidth}{!}{%
\includegraphics[width=\linewidth]{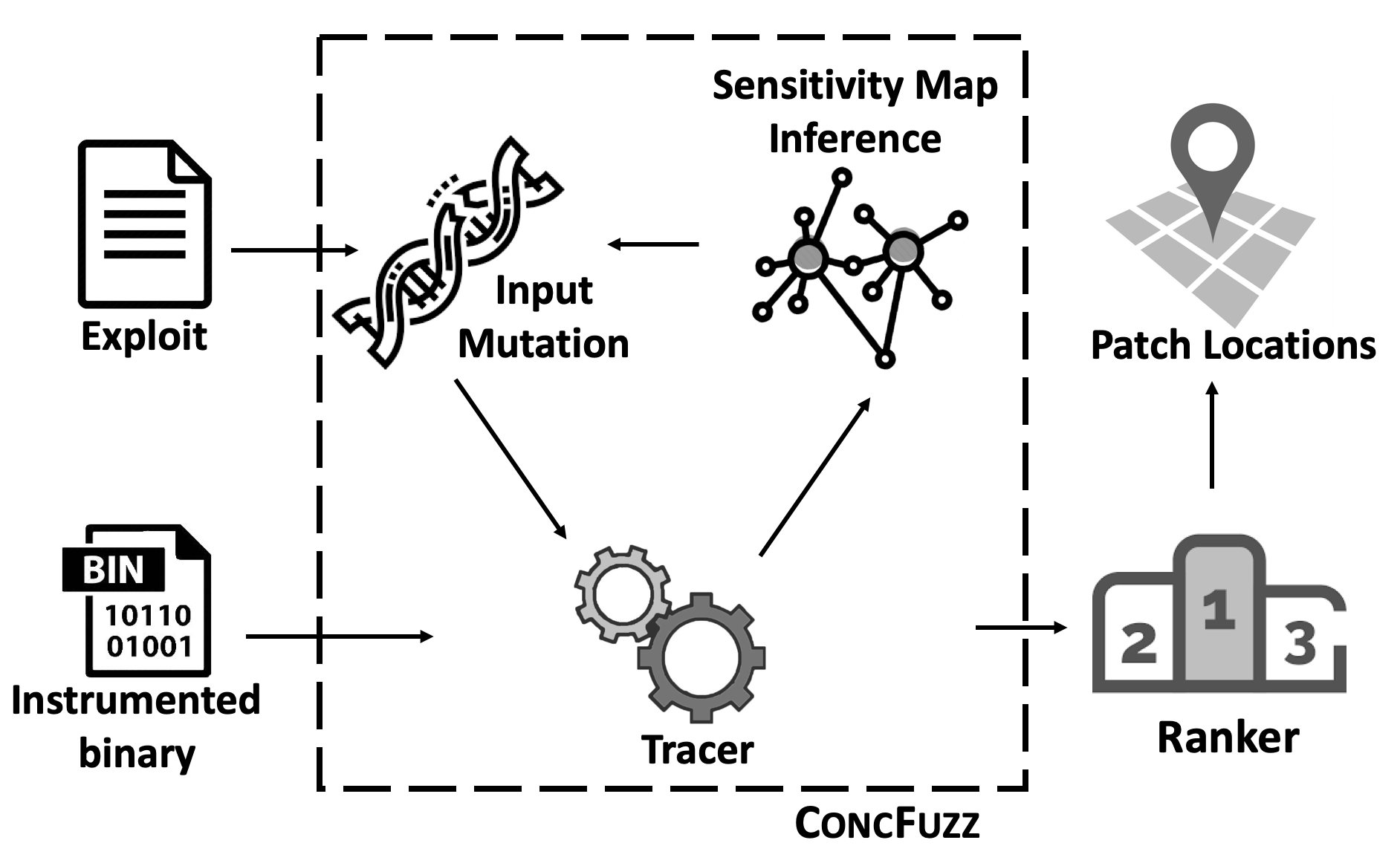}
}
\caption{
\tool's Architecture.
}
\label{fig:arch}
\end{figure}
\section{Technical Details}\label{sec:arch}

The overall architecture of our tool called \tool is shown in
Figure~\ref{fig:arch}. \tool includes two main components: the concentrated
fuzzer \fuzzer and the Ranker. \fuzzer takes in an instrumented vulnerable
program and an exploit input. It runs in a cycle until a pre-defined timeout
is reached or sufficient test cases for each $u_j$ have been generated. Given
the concentrated test-suite, the Ranker then computes the sufficiency and
necessity score and reports the Top-K location with the highest combined
score.

\subsection{\fuzzer Internals}

\begin{algorithm}[t]
\caption{\update{The implementation of concentrated fuzzing.} }
\label{alg:overall}
\begin{algorithmic}[1]
\State \textbf{Input: } Exploit input $i_e$, Instrumented Binary $Prog$
\State \textbf{Result: } Test-suite $T$
\State Seed Pool $SP \leftarrow \{i_e\}$; $T \leftarrow \emptyset$
\Repeat
	\State $i_s\leftarrow$chooseSeed($SP$)
	\State $t_s\leftarrow$execute($i_s, Prog$)
	\State Sensitivity Map $SM \leftarrow initSM(i_s, t_s)$
	\State Number of Mutated Bytes $\#B \leftarrow 0$
	\Repeat
		\State $\#B \leftarrow \#B+1$
		\Repeat
			\State Bytes $B\leftarrow$selectMutateBytes($SM, t_s, \#B$)
			\If{$B$ is empty}
				\State \textbf{Break};
			\EndIf
			\For{$j$ from $1$ to $\gamma$}
				\State $i_m\leftarrow$mutate($i_s$, $B$)
				\State $t_m\leftarrow$trace($i_m, prog$)
				\State $T\leftarrow T+\{t_m\}$
				\If{$i_m$ is an exploit with a new trace}
					\State $SP\leftarrow SP + \{i_m\}$
				\EndIf
				\State $SM\leftarrow$updateSM($SM, t_s, I_k, t_m$)
			\EndFor
		\Until{Timeout reaches or all the branch instances have sufficient test cases}
		\If{$\#B \geq \beta$ }
			\State \textbf{Break};
		\EndIf
	\Until{Timeout reaches or all the branch instances have sufficient test cases}
\Until{Timeout reaches or all the seeds are checked}
\end{algorithmic}
\end{algorithm}

Concentrated fuzzing, the high-level design of which is outlined in
Algorithm~\ref{alg:ideal}, generates a concentrated test-suite by executing
the prefix of each $u_j$. \update{The key challenge is to generate multiple inputs
which follow the prefix of $u_j$, since a random mutation of the exploit input
$i_e$ is unlikely to execute a long prefix.} To tackle this problem, the main
observation we make is that usually a small set of input bytes are responsible
for observing each $u_j$. Thus, if we can fix the values of the bytes
influencing all of $\{u_1,\cdots,u_{j-1}\}$ to their corresponding values in
the exploit trace, then any mutation made on the remaining bytes will result
in a new test case that most likely executes the prefix of $u_j$.

Precisely, let each input $i \in I$ contain $q$ bytes of which each is
represented as $b_k$ ($k \in \{1, 2, \cdots, q\}$). \fuzzer keeps track of the
{\em influence} of different inputs bytes on each $u_j$. If we can observe
that a change in the value of an input byte $b_k$ results in a change on the
state of $u_j$, we say that $b_k$ influences $u_j$, or $u_j$ is sensitive to
$b_k$.  The sensitivity relation is thus a binary relation $SM: (U, I)
\rightarrow \{0,1\}$. $SM(u_j, b_k) = 1$ if we have observed in concrete runs
that $u_j$ is sensitive to input byte $b_k$, and 0 otherwise. \update{The relation is
stored explicitly in a data structure called ``sensitivity map''.} Notice
that the sensitivity map captures both control and data dependencies between
branches and the input. This notion of influence or sensitivity is the same as
that described in recent works~\cite{chua2019one,gangreyone}, however, the way
it is used in concentrated fuzzing is quite different.
Algorithm~\ref{alg:overall} explains how \fuzzer generates a concentrated
test-suite with the help of the sensitivity map.

\paragraph{Sensitivity Map Inference.}
The sensitivity map is constructed directly from the observations during the
execution of test inputs. We mutate each byte $b_k$ of the input a constant
number of times (see $\gamma$ at Line $16$) and observe whether the state of
$u_j$ changes due to each mutation. If we cannot observe $u_j$ on the trace
after the mutation over an input byte $b_k$ in $i_e$, we infer that $b_k$
implicitly or explicitly influences the state of $u_j$ i.e., $SM(u_j, b_k)=1$
as well.

In more detail, for each round of fuzzing, \fuzzer first compares the trace
$t_s$ under a selected exploit input $i_s$ with the trace $t_m$ under each
mutated input $i_m$. If $t_m$ diverges from $t_s$ at $u_j$, \update{\fuzzer marks
$u_j$ as being sensitive to the input byte in $i_e$ mutated to create $i_m$ in the
sensitivity map.} To exemplify, let us revisit the running example from
Figure~\ref{fig:design_example}. Given the selected exploit input $i_s=(10,
15, 2)$ and the mutated input $i_m=(10, 25, 2)$---where the first byte $b_1$
has value 10, the  second byte $b_2$ is mutated and the third byte $b_3$ has
value $2$---\fuzzer infers that the branch at Line $11$ is sensitive to
$b_2$ of the input.  This is because branch at line $11$ is not observed when
$b_2>20$ in the trace of the exploit input $i_s$, and gets observed in the
trace of the mutated input $i_m$.

\fuzzer incrementally computes a sensitivity map over each mutation for each
$u_j$. It starts with an empty sensitivity map (at Line $7$) where each $u_j$
is not influenced by any input byte. Then, \fuzzer keeps updating it given a
new test case generated for each round of fuzzing (at Line $23$) until the
timeout reaches or all the branch instances are fully explored.

\begin{figure}[t]
\centering
\begin{lstlisting}[style=JavaScript, language=C, xleftmargin=0.3cm, captionpos=b]
// arr holds the fuzzer input
int buggy(char *arr){
	b = arr[0];
	c = arr[1];
	d = arr[2];
	e = arr[3];
	g = arr[4];
	ans = 0;
	...
	if (b==1 && c==4){
			...
	}
	if (d==2 || e==3){
			...
			if (g<5){
				...
				//Bug here
			}
		}
	...
	return ans;
	}
\end{lstlisting}
\caption{
See that the and condition in Line $10$ will require only single byte mutations to infer that both $arr[0]$ and $arr[1]$ influence the condition. However, the condition in Line $13$ requires you to mutate both bytes $arr[2]$ and $arr[3]$ simulataneously to infer that those bytes influence that condition.
}
\label{fig:multiple_bytes}
\end{figure}
\ash{Note that there could be scenarios where multiple bytes of the input may be
influencing the same branch and the sensitivity map inferred by single byte mutations, as described above, may not be useful for generating concentrated test-suite. For instance, consider the code in
Figure~\ref{fig:multiple_bytes}. The exploit input constitutes of $arr = [1,4,2,3,0]$ which satisfies the conditionals in Lines $10, 13$ and $15$. The first conditional statement, in Line
$10$, is a conjunction of conditionals on the values of two input bytes
$arr[0]$ and $arr[1]$. Here, mutating one input byte at a time is sufficient
to infer that the branch is sensitive to both the bytes. However, this is not
the case for the disjunction in Line $13$. In this case, single byte mutation
will lead to a conclusion that bytes $arr[2]$ and $arr[3]$ do not influence
the branch as one of them is always $True$ while the other is being mutated. Therefore, while exploring the branch $15$,our sensitivity map will show that only $arr[0], arr[1]$ as important to execute the prefix till branch $13$.  Hence, \tool would fix only the values of $arr[0], arr[1]$ and mutate other bytes including $arr[2], arr[3], arr[4]$. Consequently, the generated concentrated test-suite will have very
few traces that reach Line $15$ attributing to a small random chance that the
traces have passed condition on the prefix branch (Line $13$). Therefore, we
need to mutate both the corresponding bytes($arr[2], arr[3]$) simultaneously to infer that the
branch is sensitive to those bytes. Evidently, sensitivity map inference
is hard since the number of combinations of such multiwise mutations is exponential in
the size of input. To tackle this, \fuzzer does multiwise mutations iteratively (at Line $12$ in Algorithm~\ref{alg:overall}).  Therefore, our technique does an approximate inference; mutating one byte at a time in the first iteration and two bytes at a time in the second until a user-configured number of iterations. Further,
we show that the inferred incomplete sensitivity map is sufficient for patch
localization for a majority of CVEs.}

\paragraph{Concentrated Test-suite Generation.}
The sensitivity map helps to generate sufficiently many samples for both
events $Y_j=1$ and $Y_j=0$, having observed $u_{j-1}$. \fuzzer first extracts
all the input bytes from the sensitivity map to which $\{u_1,\cdots,u_{j-1}\}$
are sensitive. It forces these input bytes to take the same value as the
selected exploit input $i_s$. Then, to obtain enough samples for $Y_j=1$,
\fuzzer mutates the input bytes to which $u_j$ is non-sensitive in $i_s$ as
the mutation over the non-sensitive bytes likely does not change the state of
$u_j$. In contrast, for generating enough samples for $Y_j=0$, \fuzzer mutates
bytes to which $u_j$ is sensitive and keeps the non-sensitive bytes as same as
in $i_s$. For each round of fuzzing, \fuzzer selects an instance $u_j$ which
is the earliest observed on the exploit trace but does not have sufficient
test cases. Then, it follows the above approach to generate sufficiently many
test cases with a limited number of mutations (from Line $9$ to $29$).

For example, let us revisit the case of Line $11$ in
Figure~\ref{fig:design_example}. Assume the sensitivity map knows that
the branches at Line $28$ and $8$ are only sensitive to $b_3$. To
force the observation of the branch at Line $28$, \fuzzer ensures the
value of $b_3$ to be $2$ (which is as same as the exploit
input). Then, it mutates over $b_2$ and keeps the remaining bytes as
same as the exploit input to get many samples that miss the branch at
Line $11$.

\subsection{Location Ranking}\label{sec:rank}

Given the concentrated test-suite, the Ranker first removes the duplicate
traces from the test-suite to avoid biasing towards any single trace. Then, it
computes the necessity and sufficiency scores of each on-exploit location
$v_i$. It first computes these three values: 1) the number of test cases
observing $v_i$ and triggering the vulnerability ($\#(X_i=1 \land C=1)$); 2)
the number of test cases triggering the vulnerability ($\#(C=1)$); and 3) the
number of test cases observing $v_i$ ($\#(X_i=1)$). Finally, it computes the
necessity score $N=\frac{\#(X_i=1 \land C=1)}{\#(C=1)}$ and the sufficiency
score $S=\frac{\#(X_i=1 \land C=1)}{\#(X_i=1)}$.

For concreteness, we revisit our running example from
Figure~\ref{fig:design_example} and the concentrated test suite generated in
Figure~\ref{fig:example_cl}. For $v_{81}$, the sufficiency score
$P(C=1|X_{81}=1) = \frac{23}{37}$ while the necessity score $P(X_{81}=1|C=1) =
1$. Then, it normalizes both scores by \texttt{min-max scaling} and ranks
locations according to L2-norm of the normalized necessity and sufficiency
scores. The normalization function $NM$ and L2-norm score are defined as
follows:
$$
NM(N) = \frac{N-min(N)}{max(N)-min(N)}
$$
$$
NM(S) = \frac{S-min(S)}{max(S)-min(S)}
$$
$$
\text{L2-norm score} =\sqrt{NM(N)^2 +NM(S)^2}
$$
where $min(N) (min(S))$ is the minimum of the necessity (sufficiency)
score across all branch locations (similarly $max(N)$ and $max(S)$ are
defined). We use L2-norm as the ranking metric mainly because it
treats the necessity and sufficiency score equally important. Notice
that L2-norm is just one of many reasonable scoring metrics that could
be used. As the other metrics (e.g., Ochiai and Tarantula) use the
same counts as L2-norm for computing the
scores~\cite{pearson2017evaluating}, we expect them to perform
comparably on our concentrated dataset.

The Ranker reports the Top-$K$ locations as localized candidates for
patching.  If there are multiple locations with the same score, the
ranker sorts them according to the proximity to the crash point. The
closer the location to the crash location, the higher the rank.

\section{Implementation}\label{sec:impl}

We implement \tool on top of DynamoRIO~\cite{dynamorio}. We build a
DynamoRIO client to dynamically monitor the branches for each CVE. \tool is
written in Python and C++ with $1.4$K LOC.

\paragraph{Dynamic Instrumentation.}
We build a DynamoRIO client to monitor the state of each branch instance. The
client searches the specific opcodes (e.g., jle, jmp, je) which are related to
conditional statements and records the address of the conditional statement in
a file. We tried the dynamic instrumentation on both instruction and branch
level. The former could take a few minutes to hours to terminate, while the
latter takes less than a second.

\paragraph{Input Mutation - Values.} 
\tool allows users to define their input mutation strategy. The default
mutation strategy is performed on byte level which includes both single-byte
mutation and pairwise mutation. Notice that the granularity of mutation is
controlled by users. Users can define the maximum number of bytes to mutate
jointly. Furthermore, \tool allows the users to specify their strategies via a
configuration file if they have any prior knowledge about the input format.
Specifying the input format in the configuration file speeds up \fuzzer by
avoiding unnecessary mutations.

\paragraph{Input Mutation - Size.}
\ash{\tool does not support changing the size of the original input currently.
However, \fuzzer changes numeric length values and NULL-termination characters as well, implicitly changing lengths of inputs. For example, in LibTIFF, our approach will change the attribute of image length in the input file, hence changing the size of inputs.}

\paragraph{Vulnerability Oracle.}
\tool allows users to define their own oracle for detecting whether an
execution of a buggy program triggers a vulnerability.  In our evaluation, we
utilize the program crash or other detecting tools (e.g., Valgrind) as our
oracle for memory safety. For numerical errors and null dereference, we
dynamically instrument the binary with the additional checks.

\paragraph{Binary to Source mapping.}
Our entire analysis is independent of source code but we compare our Top-K
patch locations to the developer-provided patch for validating the correctness
of our results. Therefore, we implement a wrapper that maps binary
instructions to the corresponding source code statements for the convenience. 
The wrapper is built on top of \texttt{objdump} utility in Linux~\cite{objdump}.

\paragraph{Optimization: Parallelization.}
\tool uses parallelization to speed up certain tasks. In fuzzing phase, the
relationship inference is strictly sequential, however, the input mutation and
execution are independent. Thus, instead of updating the sensitivity map for
each test case, we dedicate each core to the fuzzing procedure of each
mutation target and collect the test cases. Then, we utilize the collected
test cases to update the sensitivity map once for each round of fuzzing. In
the ranking phase, the sufficiency and necessity scores for multiple locations
can be computed simultaneously before the ordering of L2-norm score.

\paragraph{Optimization: Caching.}
\tool stores the generated inputs and their corresponding traces for each
round of fuzzing. In the fuzzing phase, if \fuzzer checks all the values of an
input byte, it will avoid the mutation over the specific input byte. This
significantly increases the efficiency of \fuzzer as each execution of the
vulnerable program requires a certain amount of time.

\section{Evaluation}
\label{sec:eval}

We aim to evaluate the following research questions:
\begin{itemize}
\item {\bf [RQ1]} How effective is \tool on real-world CVEs?
\item {\bf [RQ2]} Does \fuzzer help to prevent test-suite bias and hence over-fitting?
\end{itemize}
We select a set of real-world CVEs and run \tool to generate possible patch
locations. We validate the efficacy of \tool by comparing our results to
developer patches for the CVEs as the ground truth. We extract the developer-generated
patches from the bug reports or the commits provided by the developers.

\begin{table}[t]
\centering
\caption{Vulnerable applications for evaluating \tool.}
\resizebox{0.46\textwidth}{!}{%
\begin{tabular}{|l|l|l|}
\hline
\textbf{App.} & \textbf{Description} & \textbf{LOC} \\ \hline
LibTIFF & \begin{tabular}[c]{@{}l@{}}A library for reading and manipulating TIFF \\files.\end{tabular} & $66$K \\ \hline
Binutils & \begin{tabular}[c]{@{}l@{}}A collection of tools capable of creating the \\managing binary programs.\end{tabular} & $2.7$M \\ \hline
Libxml2 & A library for parsing XML documents. & $0.2$M \\ \hline
Libjpeg & A library for handling JPEG image format. & $42$K \\ \hline
Coreutils & \begin{tabular}[c]{@{}l@{}}A collection of basic tools used on UNIX-like \\ systems.\end{tabular} & $63$K \\ \hline
JasPer & \begin{tabular}[c]{@{}l@{}}A collection of tools for coding and \\manipulating images.\end{tabular} & $28$K \\ \hline
FFmpeg & \begin{tabular}[c]{@{}l@{}}A collection of libraries and programs for \\handling video, audio and other files.\end{tabular} & $0.9$M \\ \hline
ZZIPlib & \begin{tabular}[c]{@{}l@{}}A library for extracting data from files \\archived in a single zip file.\end{tabular} & $8$K \\ \hline
Potrace & A tool for tracing bitmap images. & $9$K \\ \hline
Libming & \begin{tabular}[c]{@{}l@{}}A library for manipulating Macromedian \\Flash files.\end{tabular} & $66$K \\ \hline
Libarchive & \begin{tabular}[c]{@{}l@{}}A library which manipulates streaming \\archives in a variety of formats.\end{tabular} & $0.1$M \\ \hline
\end{tabular}
}
\label{tab:app_table}
\end{table} 

\begin{table}[!hbtp]
\centering
\caption{
Efficacy of \tool for patch localization in details [RQ1].
Column ``\#B'' shows the number of branch conditions in total. ``\#UB'' indicates
the unique on-exploit locations. 
``At Crash Loc?'' shows whether the developer-generated patch is at the crash location or not.
``Size(TS)'' means number of unique traces
generated by \tool for each CVE in $4$ hours.
 ``In Top-5?'' describes whether
 there is a correct patch location hitting one of the Top-5 candidates
 outputted by \tool. ``SM'' means that the location of the developer patch hits
 one of the Top-5 candidates. ``EQ'' indicates the existence of an
 equivalent patch located at one of the Top-5 candidates. 
 The last column
 describes the rank of \tool's output which appears in / is equivalent to developer patch
 (most have rank $\leq 5$).
}
\resizebox{0.47\textwidth}{!}{%
\begin{tabular}{|c|c|c|c|c|c|c|c|c|}
\hline
\textbf{App.} & \textbf{CVE ID} &
\textbf{\begin{tabular}[c]{@{}c@{}}Bug \\ Type \end{tabular}} &
\textbf{\begin{tabular}[c]{@{}c@{}}\#B\end{tabular}}&
\textbf{\begin{tabular}[c]{@{}c@{}}\#UB\end{tabular}} &
\textbf{\begin{tabular}[c]{@{}c@{}}At\\Crash\\Loc?\end{tabular}} &
\textbf{\begin{tabular}[c]{@{}c@{}}Size(TS)\end{tabular}} &
\textbf{\begin{tabular}[c]{@{}c@{}}In \\ Top-5?\end{tabular}} &
\textbf{Rank}\\ \hline
\multirow{17}{*}{LibTIFF}
 & CVE-2016-3186 & BO & $0.3$K & $30$ & \cmark & 14 & \cmark (SM) & $2$\\
 & CVE-2016-5314 & BO & $0.1$M & $0.7$K & \xmark &  $0.4$K & \cmark (SM) & $5$ \\
 & CVE-2016-5321 & BO & $6.6$K & $0.6$K & \cmark &  $4.3$K & \xmark & $18$ \\
 & CVE-2016-9273 & BO & $8.2$K & $0.5$K & \xmark &  $1.7$K & \cmark (EQ) & $1$ \\
 & CVE-2016-9532 & BO & $21.1$K & $0.7$K & \xmark &  $0.4$K & \cmark (EQ) & $1$ \\
 & CVE-2016-10092 & BO & $15.4$K & $0.9$K & \xmark &  $5.4$K & \cmark (EQ) & $3$ \\
 & CVE-2016-10094 & BO & $41.1$K & $1.0$K & \cmark &  $4.0$K & \cmark (SM) & $1$ \\
 & CVE-2016-10272 & BO & $1.2$M & $0.9$K & \xmark &  $19$ & \xmark & $39$ \\
 & CVE-2017-5225 & BO & $12.8$M & $0.7$K & \xmark &  $3.1$K & \cmark (EQ) & $1$ \\
 & CVE-2017-7595 & DZ & $13.1$K & $0.8$K & \xmark &  $2.7$K & \cmark (EQ) & $1$ \\
 & CVE-2017-7599 & DT & $10.2$K & $0.8$K & \cmark & $4.5$K  & \cmark (EQ) & $1$ \\
 & CVE-2017-7600 & DT & $10.3$K & $0.7$K & \cmark & $30$  & \cmark (SM) & $1$  \\
 & CVE-2017-7601 & IO & $13.5$K & $0.9$K & \cmark &  $2.4$K & \cmark (SM) & $4$ \\
 & Bugzilla-2611 & DZ & $0.1$M & $0.6$K & \xmark &  $1.4$K & \cmark (SM) & $1$ \\
 & Bugzilla-2633 & BO & $6.1$K & $0.7$K & \xmark &  $5.8$K & \cmark (EQ) & $1$ \\ \hline
\multirow{5}{*}{Binutils}
 & CVE-2017-6965 & BO & $2.3$K & $0.5$K & \cmark &  $0.4$K & \cmark (SM) & $4$ \\
 & CVE-2017-14745 & IO & $9.5$K & $0.6$K & \cmark &  $1.5$K & \cmark (SM) & $1$ \\
 & CVE-2017-15020 & BO & $16.0$K & $1.1$K & \cmark &  $1.4$K & \cmark (SM) & $1$ \\
 & CVE-2017-15025 & DZ & $28.1$K & $1.0$K & \cmark &  $1.4$K & \cmark (SM) & $1$ \\ \hline
\multirow{5}{*}{Libxml2}
 & CVE-2012-5134 & BO & $8.1$K & $1.5$K & \cmark &  $43.2$K & \cmark (SM) & $1$ \\
 & CVE-2016-1838 & BO & $0.4$M & $1.0$K & \cmark &  $4.7$K & \cmark (SM) & $1$ \\
 & CVE-2016-1839 & BO & $1.5$M & $1.4$K & \xmark &  $0.9$K & \cmark (SM) & $1$ \\
 & CVE-2017-5969 & ND & $22.5$K & $1.4$K & \cmark &  $10.0$K & \xmark & $24$ \\ \hline
\multirow{4}{*}{Libjpeg}
 & CVE-2012-2806 & BO & $1.5$K & $0.2$K & \cmark &  $46$ & \cmark (SM) & $1$ \\
 & CVE-2017-15232 & ND & $0.1$M & $0.6$K & \cmark &  $8.0$K & \cmark (SM) & $1$ \\
 & CVE-2018-14498 & BO & $1.2$K & $0.1$K & \cmark &  $0.1$K & \cmark (SM) & $1$ \\
 & CVE-2018-19664 & BO & $21.3$M & $0.1$K & \xmark &  $5$ & \cmark (EQ) & $3$ \\ \hline
\multirow{4}{*}{Coreutils}
 & GNUbug-19784 & BO & $0.2$K & $34$ & \cmark &  $0.5$K & \cmark (SM) & $1$ \\
 & GNUbug-25003 & IO & $0.1$K & $0.1$K & \cmark &  $7$ & \cmark (SM) & $1$ \\
 & GNUbug-25023 & BO & $1.3$K & $0.3$K & \xmark &  $0.1$K & \xmark      & $>200$ \\
 & GNUbug-26545 & IO & $0.5$K & $0.2$K & \xmark &  $1.9$K & \cmark (SM) & $2$ \\ \hline
\multirow{3}{*}{JasPer}
 & CVE-2016-8691 & DZ & $38.9$K & $0.3$K & \xmark &  $0.1$K & \cmark (EQ) & $1$\\
 & CVE-2016-9557 & IO & $44.0$K & $0.5$K & \cmark &  $2.7$K & \cmark (SM) & $4$\\ \hline
\multirow{2}{*}{FFmpeg}
 & CVE-2017-9992 & BO & $11.7$K & $0.6$K & \cmark &  $0.6$K & \cmark (EQ) & $1$\\
 & Bugchrom-1404 & IO & $7.6$M & $0.9$K & \xmark &  $0.4$K & \xmark & $187$\\ \hline
\multirow{3}{*}{ZZIPlib}
 & CVE-2017-5974 & BO & $0.1$K & $0.1$K & \xmark &  $0.2$K & \cmark (SM) & $2$ \\
 & CVE-2017-5975 & BO & $0.1$K & $0.1$K & \xmark &  $0.2$K & \cmark (EQ) & $2$ \\
 & CVE-2017-5976 & BO & $0.1$K & $0.1$K & \cmark &  $0.3$K & \cmark (SM) & $1$ \\ \hline
Potrace
 & CVE-2013-7437 & BO & $0.3$M & $0.1$K & \xmark &  $2$ & \cmark (EQ) & $1$ \\ \hline
 \multirow{3}{*}{Libming}
 & CVE-2016-9264 & BO & $38$ & $26$ & \xmark &  $29$ & \cmark (EQ) & $4$ \\ 
 & CVE-2018-8806 & UF & $1.1$K & $0.1$K & \cmark & $2.3$K  & \cmark (EQ) & $2$ \\
 & CVE-2018-8964 & UF & $1.1$K & $0.1$K & \cmark & $4.6$K  & \cmark (EQ) & $5$ \\ \hline
Libarchive
 & CVE-2016-5844 & IO & $6.1$K & $0.7$K & \cmark &  $46$ & \cmark (SM) & $1$ \\ \hline
\end{tabular}
}

\label{tab:cve_result}
\end{table} 

\subsection{Subjects and Setup}

Our subjects are chosen to satisfy three requirements: 1) The vulnerable
applications can be executed; 2) a working exploit is available; and 3) a
valid developer patch is available.

\paragraph{Diversity of Subjects}
\update{%
We select \fuzznum CVEs that correspond to \fuzzappnum applications, shown in
Table~\ref{tab:app_table}. Our dataset includes all $15$ CVEs from the
existing benchmarks used by recent work~\cite{senx} that satisfy the above
three criteria\footnote{SENX has $42$ benchmark programs. We eliminated the
following: $18$ programs that do not have any developer patches (missing
ground truth to evaluate against); $2$ that do not have reproducible exploits,
$2$ that are on x86 CPUs while our present implementation supports only x64;
$5$ that do not work on vanilla DynamoRio without instrumentation (either
crashing DynamoRio or taking hours and utilizing excessive memory for a single
trace).}.
We added $28$ more CVEs to increase the diversity of the benchmarks, as SENX
benchmarks have only $2$ kinds of security vulnerabilities. Our final
benchmarks have $6$ categories of vulnerability including $26$ buffer overflow
(BO), $4$ divide-by-zero (DZ), $7$ integer overflow (IO), $2$ null pointer
dereferences (ND), $2$ heap use-after-free (UF) and $2$ data-type overflows (DT).}

\paragraph{Statistics of Subjects.}
The subjects have sizes ranging from $10$ thousand to $2$ million LOC.
Most of them  have very few (less than $30$) or no manually written tests for
the vulnerable program and its configuration. The exploit input sizes vary from
$1$B to $74$KB with an average of $8$KB. Table~\ref{tab:cve_result} shows that
the exploit traces have a few tens to millions of observed on-exploit branch
instances. Recall that \tool works by recording only branch conditionals,
i.e., one per basic block. On average, there are $0.3$ million on-exploit
branch instances, with minimum of $38$ and maximum of $7.6$ million.  Due to
loops and recursion, many observed locations repeat---we also report the
unique number of locations considered by \tool for computing scores in
Table~\ref{tab:cve_result}.

\paragraph{Experiment Setup.}
All our experiments are performed on a $56$-core $2.0$GHz $64$GB RAM Intel
Xeon machine. Each round of fuzzing phase allows to mutate maximally $2$ bytes at a time
($\beta$=$2$ in Algorithm~\ref{alg:overall}) and mutates $\gamma$ times over
each mutation target. We set $\gamma$=$200$ (see  Algorithm~\ref{alg:overall}) for the default mutation strategy. We set a timeout of $4$ hours
per benchmark to generate a test-suite and allow the fuzzing phase to fork
$10$ processes maximally.

\paragraph{Correctness criteria.}
We say that \tool is able to pinpoint the fix location, if:
\begin{itemize}
\item the Top-5 locations outputted by \tool includes (at least one) location of the developer patch; or
\item one of the Top-5 locations from \tool can be modified to produce a patch semantically equivalent to developer patch.
\end{itemize}
\ash{The semantically equivalent patch is generated based on the distance of predicted location to the developer provided patch location and available live variables. If the predicted location is in the same function as the developer patch and all the variables used in the developer patch are live, a simple displacement of developer's patch usually suffices. If that is not possible then we use domain specific knowledge to create a patch using the developer patch variables or the live variables that taint them. In order to validate our patches we run the patched application on the generated test-suite as well as the developer provided test suite if available. An example of equivalent patch is described in Section~\ref{sec:case_study}. For more instances, we also provide all the equivalent patches in the supplementary material.}
Our criterion of choosing Top-5 recommendations follows from empirical
studies on practitioners' expectations from automated fault
localization tools~\cite{kochhar2016practitioners}. We also report on
the exact rank of the correct patch in Table~\ref{tab:cve_result}.

\subsection{[RQ1] Efficacy for Patch Localization}

 \begin{figure}[t]
 \centering
 \resizebox{0.45\textwidth}{!}{%
 \includegraphics[width=\linewidth]{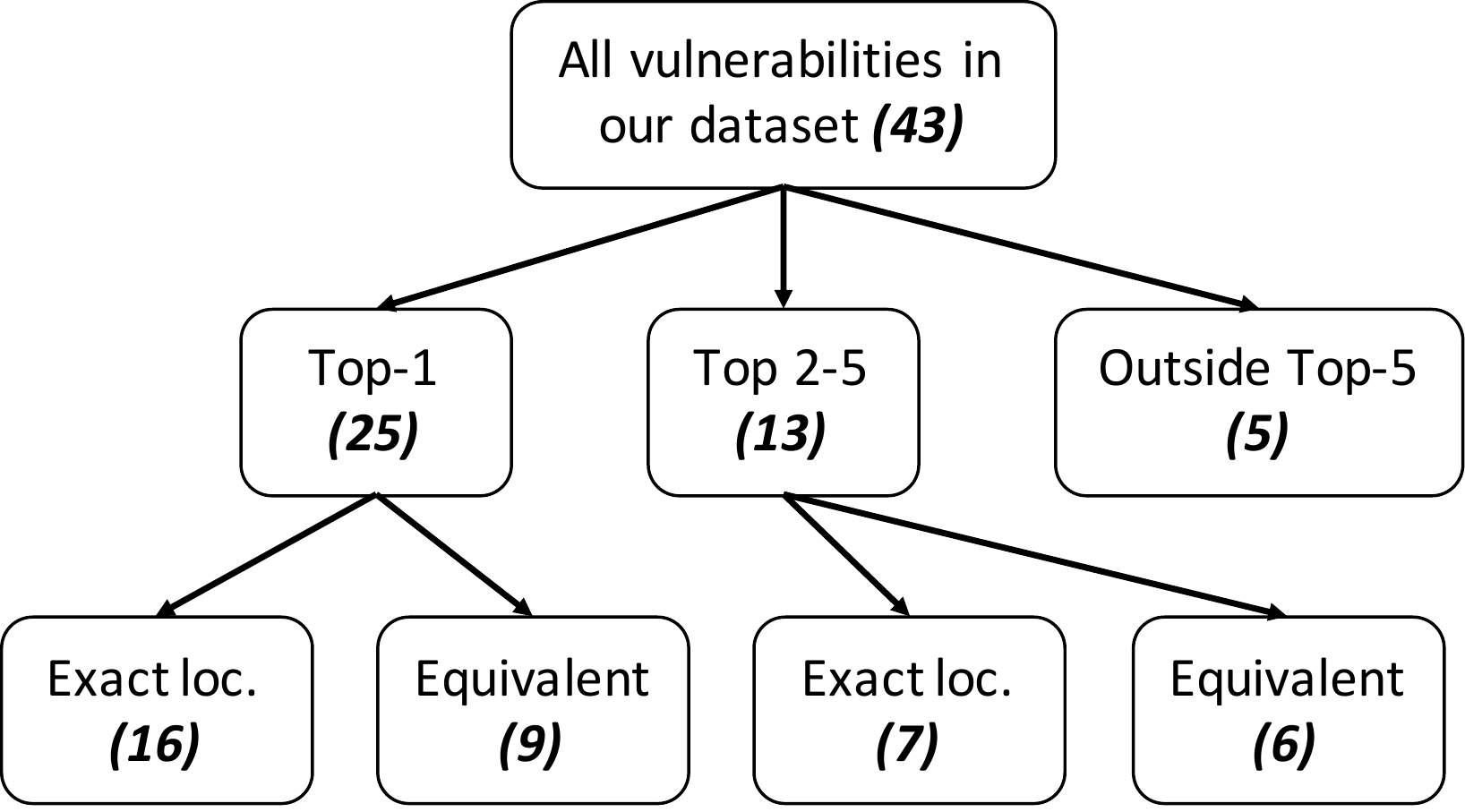}
 }
 \caption{Efficacy of \tool for patch localization [RQ1].}
 \label{tab:cve_summary}
 \end{figure}

\begin{figure*}[t]
\center
\resizebox{0.8\textwidth}{!}{%
\begin{tikzpicture}
\begin{axis}[
    ybar,
    width=1.0\textwidth,
    height=.25\textwidth,
    xlabel={Range of Distinguishability Ratio},
    ylabel={\#(CVE)},
    xtick=data,
    xticklabels={{$[0,0.1]$}, {$(0.1,0.2]$}, {$(0.2,0.3]$}, {$(0.3,0.4]$}, {$(0.4,0.5]$},
                 {$(0.5,0.6]$}, {$(0.6,0.7]$}, {$(0.7,0.8]$}, {$(0.8,0.9]$}, {$(0.9,1.0]$}},
    legend pos=north west,
    ymajorgrids=true,
    grid style=dashed,
]

\addplot[
    postaction={pattern=dots},
    ]
    coordinates {
    (1, 3) (2, 7) (3, 11) (4, 10) (5, 5) (6, 4) (7, 0) (8, 0) (9, 0) (10, 3)
    };\label{plot1}
\addlegendentry{T1}

\addplot[
    postaction={pattern=north east lines},
    ]
    coordinates {
    (1, 3) (2, 12) (3, 12) (4, 4) (5, 5) (6, 0) (7, 2) (8, 2) (9, 0) (10, 3)
    };\label{plot2}
\addlegendentry{T2}

\end{axis}
\end{tikzpicture}
}
\caption{Effects of test-suite bias [RQ2].}
\label{fig:overfit_summary}
\end{figure*}
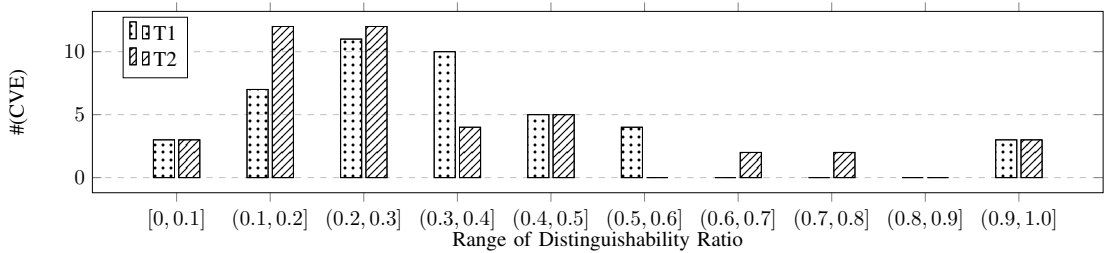

\paragraph{Main Results.}
Figure~\ref{tab:cve_summary} summarizes the efficacy of \tool for patch
localization and the distribution of the type of the generated patch location.
\shiqi{Out of \fuzznum CVEs, \tool successfully locates the patch for \patchnum CVEs
within the Top-5 candidates, Among these \patchnum CVEs, the patch location
for $25$ CVEs hits the topmost candidate (see Figure \ref{tab:cve_summary}).
Recall that there may exist multiple patch locations which are equivalent for
fixing the vulnerability. We observe that, for $23$ out of \patchnum successful
CVEs, one of the top-5 candidate locations corresponds exactly to a location
patched in the developer patch. For $15$ out of \patchnum CVEs, we can create
an equivalent patch.}  To further investigate these results,
Table~\ref{tab:cve_result} presents the detailed result of patch localization
for each CVE. \tool successfully generates sufficient test-suite for
each CVE. Unlike the manually written test-suite where no test triggers the
bug, \tool generates $2.7K$ test cases on average, around $40$\% of which trigger
the vulnerability for half of the benchmarks. \shiqi{In addition, \tool performs
well on all categories of security bugs: it successfully locates the patch in
Top-5 candidates for $23$ buffer overflows, $6$ integer overflows, all $4$
divide-by-zero, $1$ null dereference, $2$ heap use-after-free and $2$ data-type overflows.} In addition, \tool performs equally
on different applications. \shiqi{For example, it successfully locates the patch for
$13$ out of $15$ CVEs in \texttt{LibTiff}, all $4$ CVEs in \texttt{Binutils},
and $3$ out of $4$ CVEs in \texttt{Libxml2}.} This indicates that the success
of \tool is not correlated with the size and the type of the application.

\paragraph{Performance.}
The total time taken for patch localization on each CVE has two components:
fuzzing time and analysis time. We set the fuzzing time to $4$ hours for all
the CVEs. The analysis time varies with each CVE and the number of candidates
to report (e.g., Top-$100$). The maximum analysis time taken by \tool is
within $10$ minutes with Top-$200$ candidates to report.

\paragraph{Distance to Crash Locations.}
One way of localizing patches is to place them right before the crash
point~\cite{senx}. \shiqi{In around $44$\% of the CVEs we study, the developer-generated patches
do not coincide with the crash location.} Our patches created from the \tool
are on a location different from the crash location for \notcrashloc out of
\patchnum CVEs. \shiqi{An example of such a patch generated by \tool, which is far
from the crash location, is CVE-2016-5314, as shown in Section~\ref{sec:prob}.}

\paragraph{Need for Probabilistic Approaches.}
In many CVEs (\lessthanone out of \fuzznum) the patch locations do not have
both necessity and sufficiency scores equal to $1$, even for developer
patches. Such patch locations do {\em not} cleanly separate all exploiting
test cases from benign ones. The lack of any program points, at which a clean
separation between passing and failing test is possible, highlights the
inherent uncertainty in choosing between patch candidates. This motivates the
need for probabilistic approaches such as ours.

\subsection{[RQ2] Tackling Over-fitting}\label{sec:overfit}

We examine the impact of the test-suite bias on patch localization.
Poor test-suites make it difficult to distinguish between different
program locations as patch points. This is exhibited by many locations
obtaining the same score from localization. On the other hand, a
concentrated test-suites segregates locations better. We can therefore
measure how much a test-suite contributes towards segregating patch
locations in statistical localization.

We evaluate patch localization using three different kinds of
test-suites:
\begin{itemize}
	\item T1: a biased test-suite which only contains  exploits,
	\item T2: a biased test-suite which only contains tests reaching the crash location, and
	\item T3: a concentrated test-suite produced by \tool.
\end{itemize}
We measure the efficacy of \tool under these three test-suites by counting the
  number of branch locations which have the same score. We call a set of
 locations with the same score as a cluster. Notice that if the test-suite is
effective in patch localization, the number of clusters will be very large. To
measure the distinguishability of a given test-suite, we set T3 as the
baseline and compute the ratio of the number of clusters generated by using T1
or T2 vs. the number of clusters using T3, which is called as 
\texttt{distinguishability ratio}.

Figure~\ref{fig:overfit_summary} summarizes the distinguishability ratio of
the biased test-suites T1 and T2 on patch localization for \fuzznum real CVEs.
\shiqi{For $36$ out of \fuzznum CVEs ($84$\%), the number of clusters generated by
T1 is 50\% fewer than the one generated by the concentrated test-suite T3.
Similar results (50\% fewer clusters for $36$ CVEs) are also shown by the biased
test-suite T2.}  This clearly demonstrates that a concentrated test-suite (T3)
improves significantly over other test-suites. 

Existing tools like AFLGo~\cite{AFLGo} and F3~\cite{jin2013f3} can be
used to generate the test-suites for our purpose. However, they are {\em not}
designed to produce concentrated test-suites, which is the key conceptual
advance in our proposed technique (see Section~\ref{sec:approach}). We
experimentally show both F3 and AFLGo generate test-suites which are biased
towards the crash location, thus, their test-suites belong to the category
$T_2$. Furthermore these tools rely on external source-based analysis engines
such as dynamic symbolic analysis (for F3) and intra-procedural control flow
graph construction (for AFLGo). 

\paragraph{Comparison with AFLGo.}
We compared our work quantitatively with the directed fuzzer AFLGo
\cite{AFLGo}. We collect all the inputs generated by AFLGo as our test-suite,
with the crash location as the target. These inputs include the both cases
which reach and deviate from the crash location. \shiqi{Although this test-suite is
balanced to some extent (and hence helps AFLGo), AFLGo can only successfully
locate the patch location in Top-5 for $18$ out of \fuzznum CVEs; this is also
somewhat because of the complexity of partial control flow graph construction
in AFLGo. In comparison, our approach indicated the patch location among Top-5
candidates in \patchnum out of \fuzznum CVEs in total. }These results show that while
our concentrated fuzzing is a form of directed fuzzing, directed fuzzing tools
cannot be straightforwardly used for our problem.

\paragraph{Comparison with F3.}
We also compare with the fault localization tool
F3~\cite{jin2013f3}. We keep the same ranking algorithm used in \tool
and only change the test-suite for a fair comparison over the quality
of the test-suite. The implementation of F3 uses an out-of-date LLVM
version, 2.9. Due to insufficient support of external functions and
the inline assembly functions, F3 fails to generate test-suite for
$19$ CVEs. We do not know how F3 would have performed in localization
accuracy for these 19 CVEs if the tool implementation was able to
handle them.  For the remaining $\fuzznum - 19 =24$ CVEs, the size of the
test-suite generated by F3 is around $4$ times smaller than the
test-suite generated by \tool. In our experiments, F3 always
recommends patch locations at or next to the crash locations. The
reason is overfitting: The test-suite obtained from F3 has a high
density of tests that reach the crash point.
If a given CVE (e.g., CVE-2016-9264 in Section~\ref{sec:case_study})
cannot be patched before the crash location, F3 fails to pinpoint the
correct patch location within Top-5 candidates. \shiqi{Among the $24$ CVEs
that F3 handles, it generates patch location in Top-5 for $19$ out of
them. In contrast, \tool generates patch location among Top-5 for all the CVEs where F3 works and $3$ more (total $22$).}

\section{Case Studies}\label{sec:case_study}

In order to understand the quality of patch localization, we present two
examples: a) CVE-2016-3186 for which the developer patch coincides with one of
the Top-5 candidates and b) CVE-2016-8691 for which the developer patch does not
coincide with any of the Top-5 candidates but there is an {\em equivalent}
manually generated patch at one of the Top-5 candidates.

\begin{figure}[t]
	\begin{lstlisting}[style=JavaScript, language=C, xleftmargin=0.3cm, captionpos=b]
		int readextension(void){
		   ...
		   char buf[255];
		   ...
		- while ((count = getc(infile)) && count <= 255)
		+ while ((count = getc(infile)) && count >= 0 && count <= 255)
		     if (fread(buf, 1, count, infile) != (size_t) count) {...}
		}
	\end{lstlisting}
	\caption{\tool highlights the same location where the developer-generated patch is applied for CVE-2016-3186.}
	\label{fig:20165321}
\end{figure}

\paragraph{Finding developer-generated patch location (CVE-2016-3186)}
This is a buffer overflow in \texttt{LibTIFF} which causes a denial of service
via a crafted GIF image. Consider Figure~\ref{fig:20165321}, the overflow
happens in function \texttt{readextension} when it reads a GIF extension block
at Line $7$. When \texttt{getc} detects the end of file, it returns $EOF$
which is negative number. However, the loop condition only checks if
\texttt{count} $\leq 255$. If \texttt{count} is negative, the loop condition
is satisfied and \texttt{count} is casted to \texttt{size\_t}, which leads to
the buffer overflow. \tool analyzes this CVE and outputs the branch condition
in Line $5$ as one among the Top-5 candidates. This coincides exactly with the
developer patch which adds an additional check at Line $5$ to prevent
overflow.

\begin{figure}[t]

	\begin{lstlisting}[style=JavaScript, language=C, xleftmargin=0.3cm, captionpos=b]
	   ...
	   samplerate_idx = (flags & MP3_SAMPLERATE) >> MP3_SAMPLERATE_SHIFT;
	+  if (samplerate_idx < 0 || samplerate_idx > MP3_SAMPLERATE_IDX_MAX){
	+     error("invalid samplerate index");}
	   ... // <-- code with no relation to samplerate_idx
	   samplerate = mp1_samplerate_table[samplerate_idx];
	\end{lstlisting}
	\caption{Developer-generated patch for CVE-2016-9264.}
	\label{fig:dev}
\end{figure}

\begin{figure}[t]
	\begin{lstlisting}[style=JavaScript, language=C, xleftmargin=0.3cm, captionpos=b]
	   ...
	   samplerate_idx = (flags & MP3_SAMPLERATE) >> MP3_SAMPLERATE_SHIFT;
	   ... // <-- code with no relation to samplerate_idx
	+   if (samplerate_idx < 0 || samplerate_idx > MP3_SAMPLERATE_IDX_MAX){
	+     error("invalid samplerate index");}
	   samplerate = mp1_samplerate_table[samplerate_idx];
	\end{lstlisting}
	\caption{Semantically equivalent manually generated patch at the location highlighted by \tool for CVE-2016-9264.}
	\label{fig:man}
\end{figure}

\paragraph{Finding equivalent patch location (CVE-2016-9264)}
This is an example of an out-of-bounds read in \texttt{Libming} library which
can crash any web application that uses this library to process untrusted mp3
files. Consider Figure~\ref{fig:dev}, the variable \texttt{samplerate\_idx} in
Line $2$, is read from an input mp3 file and is used to set the
\texttt{samplerate} in Line $6$. Executing the exploit mp3 file results in an
out-of-bounds access at Line $6$ which sets \texttt{samplerate} to $0$ and
later results in a crash due to floating-point exception. So, the developer
patch is applied at Line $3$ just after reading \texttt{samplerate\_idx} from
input. However, \tool suggests to add a check just before the out-of-bounds
access at Line $6$, shown in Figure~\ref{fig:man}. \update{The original code 
between Line $2$ and Line $6$ does not use \texttt{samplerate\_idx} and it is
not affected by the input file.}

\section{Discussion}

\paragraph{Ruling out spurious correlation.}
Correlation does {\em not} imply causation, and given the statistical
nature of \codename, it is natural to ask whether the results observed
are an artifact of pure chance or spurious correlations. We
additionally investigated why \codename works in the cases where it 
reports the right candidate in the Top-5.

First, we observed that the correct developer-provided patch is small,
typically spanning a single branch location or at most $3$ branch locations for
more than $90$\% of our benchmarks.  Given that each benchmark executes
thousands of basic blocks in one exploit, the odds of pinpointing the correct
branch location in the Top-5 by random chance is extremely low. \codename is
doing significantly better than randomly guessing locations.

Second, we manually investigated why \codename assigns the highest
score to the correct patch location whenever it does.  To carry out
this investigation, we extended \codename to compute the sensitivity
map for the variables around that location. Upon testing with the
concentrated test-suite, we found that certain variables have the
highest L2 scores---they are most sensitive to transformation of a
benign input into an exploiting one. We find these highest sensitivity
variables often correspond to the variables that are sanitized or
bounded in the developer-provided patch. For example, the variable
$count$ has been correctly identified as the highest sensitivity
variable for \texttt{CVE-2016-3186}. Our manual investigation confirms
that a simple extension to \codename is able to identify a handful of
candidate variables that should be patched, beyond just identifying
the correct location. This shows that the results that \codename is
explainable and not an artifact of spurious correlation.
We leave utilizing this observation for a full patch synthesis to
future work.

\paragraph{Quality of Patch Locations.}
During our manual analysis over patch locations, two of the authors
independently analyzed the location of developer-generated patch
versus the location recommended by \tool. In particular, for the
equivalent patch location, as the developer patch is available,
generating a semantically equivalent patch and inspecting it manually
turned out to be relatively straightforward in our experiments,
requiring less than an hour of work per CVE per person.  

\paragraph{Sensitivity map recovery.} 
The recovery of the sensitivity map may be of independent interest to
other binary analyses and fuzzers. \codename uses a simplistic
strategy to recover the sensitivity map in its \fuzzer module.  It
uses single byte mutations as well as pairwise mutations. We observed
that using pairwise mutations improves the recovered sensitivity map
over using single byte mutations in our experiments, at the expense of
increasing the number the number of tests quadratically. We believe
that more advanced strategies could be employed, for example, based on
group mutations combined with binary search. This can further improve
the scalability of \codename or other tools that use sensitivity maps.

\paragraph{Evaluation Subjects.} 
To mitigate risks of selection bias, we chose application subjects /
CVEs from a recent study on security bug repair \cite{senx}. To show
the generalization of \tool over various bug types, we add $24$ more
CVEs into our evaluation subjects with four bug types in total.
However, different benchmarks may lead to different results---this
remains a threat to validity for our work.

\paragraph{Multi-line Patches.}
\ash{PatchLoc currently does not support multiple locations for patching. We speculate that multi-location patches may be feasible in future work by calculating necessity / sufficiency scores for multiple locations i.e., considering the joint distribution across multiple locations at a time.}

\section{Related Work}

One of the earliest efforts in fault localization is via dynamic slicing
\cite{AgrawalHorgan1990}. It takes in a program input and a slicing criterion
in the form of $\langle l, v \rangle$ where $l$ is a location and $v$ is a
variable. It uses data and control dependencies to explain the value of $v$ in
$l$ in the execution trace of the given input.  Since dynamic slicing involves
high computational overheads and dynamic slices are still large, more accurate
methods to localize observable errors in programs have been studied.  One of
the notable works in this regard, is the principle of delta
debugging~\cite{ZellerTSE} which localizes observable errors by computing the
differential of a failing artifact, and a ``similar" benign artifact. The
artifact could be in the form of test inputs, or execution traces.  One of the
major difficulties in employing this line of work is that its accuracy
crucially depends on the choice of the benign artifact.

Progress in localization via trace comparison has led to other works involving
more systematic generation of the benign trace, and a natural extension to
probabilistic reasoning. These include the use of a systematic off-line search
to generate the passing trace via branch direction mutation \cite{WangASE05},
as well as online predicate switching by forcibly switching a branch
predicate's outcome at run-time \cite{ZhangPLDI06}. Our work draws some
inspiration from the theme of predicate switching, however, it is effected in a
completely different fashion. Instead of forcibly changing a branch predicate
at run-time, we conduct repeated runs of directed fuzzing with the goal of
flipping branch predicate(s). 

Our work follows the statistical fault localization framework~\cite{Survey},
where a score is assigned to each statement of the program based on its
occurrence in passing and failing execution traces. One of the first works in
this regard is  Tarantula~\cite{TarantulaICSE02}, which has subsequently been
followed by many works proposing many scoring metrics, including the Ochiai
metric \cite{OchiaiAbreu}. The main hypothesis in these works is that the
control flow of the execution traces of tests can be used to determine likely
causes of failure of a test. Thus, if a statement occurs frequently in failing
test executions and rather infrequently in passing test executions, it is
likely to be scored highly and brought to the attention of the developer. It
is well-known that the accuracy of these methods is highly sensitive to the
choice of tests~\cite{pacheco2007feedback,tillmann2008pex}. Most works in this
regime use externally provided or arbitrarily chosen test suites.

Very few works have attempted to address the central challenge of choosing the
right test suite. Works related to ours include F3~\cite{jin2013f3} which
builds on the techniques proposed in BugRedux~\cite{jin2012bugredux}. The goal
of BugRedux is different from ours, it is to re-produce a field failure trace
by following through "breadcrumbs" given as locations visited.  F3
\cite{jin2013f3} relaxes the execution synthesis component of BugRedux by
generating many tests via symbolic execution. Hence F3 is closer to our work
than BugRedux and we can compare our concentrated fuzzing with test generation
in F3. Our quantitative comparison with F3 has been reported in this paper.
Other works like MIMIC \cite{zuddas2014mimic} extend F3 with a model of
correct behavior developed from dynamic specification mining in the form of
potential invariants from passing traces. Such works are geared towards
explaining failure causes for better debugging, whereas we identify locations
for inserting patches.

\ash{An independent and concurrent work called AURORA also proposes
  patch localization under similar assumptions, however, it uses an
  off-the-shelf fuzzing strategy, namely AFL's crash exploration mode,
  to create a test-suite for statistical fault
  localization~\cite{aurora2020usenix}. AURORA proposes mechanisms for
  synthesizing and ranking a particular kind of predicates during its
  statistical analysis. In contrast, \tool offers a new systematic
  test-suite generation technique, while retaining the rest of the
  structure of statistical fault localization. We believe that our
  work is complementary as one could combine our concentrated
  test-suite generation with the predicate synthesis and ranking mechanism
  proposed in AURORA.}

Even though \tool does not synthesize patches, the task of patch localization
via concentrated fuzzing can be seen as a mechanism to alleviate over-fitting
in program repair. Compared to existing works which heuristically rank
candidate patches to reduce over-fitting \cite{patchrank1,patchrank2}, \tool
supports systematic test generation to witness the possible deviations from a
given exploit trace. 

A different line of work employs symbolic analysis methods for localizing the
root cause of an observable error
\cite{DARWIN09,MajumdarPLDI11,adebug,WiesFM12}. The central observation in
these works is that localization can benefit from specification inference. 
Even in the absence of formal specifications of intended program behavior,
these works seek to infer properties of intended program behavior by
symbolically analyzing various program artifacts such as failing execution
traces, past program versions as so on. These approaches proceed via source
code analysis, and incur the overheads of symbolic execution.

Our specific proposal for concentrated fuzzing is most closely related to
GREYONE, a recent work of taint-based fuzzing for
bug-finding~\cite{gangreyone} which extends notions of taint or influence from
recent work~\cite{chua2019one}. Concentrated fuzzing has orthogonal objectives
to this work, as it does not aim to maximize coverage or number of exploits.

SENX is an automatic patch synthesis tool for certain vulnerabilities based on
information from source code and an exploit~\cite{senx}. SENX uses a
simplistic strategy for localization: it uses the statement before the crash
as the patch point. Such localization is typically only sufficient for
if-guard fixes at the crash location, which may not fix the fault in a general
way, but workaround to prevent an error from being observable. In our
experiments, we have reported 10 (out of 34) legitimate patch locations which
are different from the crash location. We show an example in Section
\ref{sec:case_study}.

Other works that aim to localize by identifying workarounds that make errors
unobservable have also been proposed, such as Talos~\cite{huang2016talos}.
Talos extensively uses source code and specializes for specific software
coding practices or idioms. A number of prior works use source code for patch
localization, including a recent work that employs deep learning over code
features~\cite{li2019deepfl}.  Our work minimizes assumptions about the
availability of such features and yet achieves high accuracy in real-world
programs.

\section{Conclusion}

In this paper, we propose a novel directed fuzzing approach to generate a
concentrated test-suite for ranking potential patch locations for an
exploitable vulnerability witnessed by a given exploit trace. We have shown
that \tool achieves high accuracy in identifying the right patch location for
real-world CVEs. From the point of view of localization, our proposed approach
can be seen as a fine-grained localization method---given an exploit trace
(essentially a trace of a failing input), we seek to systematically generate
tests whose execution summaries can provide an explanation of the failure by
suggesting candidate patch locations.

In conclusion, we highlight four important aspects of our technique. First it
does not depend on manually constructed tests, and systematically generates
deviations from an exploit via a form of controlled fuzzing.  Second, and
related to the first point, we achieve the systematic test-suite construction
without incurring the overheads of  symbolic approaches such as symbolic
execution. Third, our approach works on both source code and binaries.  So, it
can work on applications where part of the source code (say of a library) is
unavailable. It can also help create a find-and-fix cycle where we seek to fix
the vulnerabilities found through fuzz testing. Last but not the least, given
the fix location(s) produced by \tool, the observed values at the fix
location(s) on the test inputs generated by concentrated fuzzing, can be
directly used as specifications to drive program synthesis engine. Such
automatic fix synthesis remains an attractive direction of future work.

\section{Acknowledgments}
We thank Shruti Tople, Shweta Shinde, Shin Hwei Tan, Teodora Baluta, Ahmad Soltani and
the anonymous reviewers for helpful feedback on this work. We
thank Jinsheng Ba for helping us in experiments. All opinions expressed in this
paper are solely those of the authors. This research is supported
by research grant DSOCL17019 from DSO, Singapore.

\balance

\bibliographystyle{IEEEtranS}
\bibliography{paper}

\end{document}